\documentclass{article}
\usepackage{amsfonts}
\usepackage{graphicx}
\usepackage{amsmath}

\input{tcilatex}

\begin{document}

\title{Nondemolition observation of a free quantum particle}
\author{V. P. Belavkin\thanks{%
M.I.E.M., D. Vusovsky 3/12, Moscow 109028, USSR} \ and P. Staszewski\thanks{%
Institute of Physics, N. Copernicus University, Toru\'{n}, Poland}}
\date{28 May 1991\\
Published in: Phys. Rev. A, \textbf{45} (1992) No 3, 1347--1356}
\maketitle

\begin{abstract}
A stochastic model of a continuous nondemolition observation of a free
quantum Brownian motion is presented. The nonlinear stochastic wave equation
describing the posterior dynamics of the observed quantum system is solved
in a Gaussian case for a free particle of mass $m>0$. It is shown that the
dispersion of the wave packet does not increase to infinity like for the
free unobserved particle but tends to the finite limit $\tau_{\infty}^2=(%
\hbar/2\lambda m)^{1/2}$ where $\lambda$ is the accuracy coefficient of an
indirect nondemolition measurement of the particle position, and $\hbar$ is
Planck constant.
\end{abstract}

\section{Introduction}

The Schr\"{o}dinger equation describes the time--development of the wave
function of a quantum system only for the time intervals between the
succeeding instants of measurements. At the instant of a measurement of some
observable with a discrete spectrum, $Z$, the quantum system makes an
immediate transition (jump) from the state $\psi (t)$ to the eigenstate $%
\psi _{z}(t)$ corresponding to the obtained eigenvalue $z$ of $Z$ with the
probability $|\langle \psi (t)|\psi _{z}(t)\rangle |^{2}$. Such a stochastic
time--behaviour of the system at the instant of the measurement assures the
repeatability of the results of measurements, if a second measurement were
taken immediately after the first one then for discrete observable $Z$ the
measurement would again give $z$ [1]. It is intuitively obvious that if one
would perform measurements with a high frequency -- in a limit continuously
in time -- the quantum system would show a stochastic irreversible behaviour
for the whole period of observation. Therefore the time--development of a
continuously observed quantum system cannot be governed by the deterministic
Schr\"{o}dinger equation describing the reversible motion. This statement
remains true also in the case of the measurements of an observable with a
continuous spectrum though for observables with continuous spectra the
repeatability hypothesis is not assumed [1--4] as, in general, there are
non--zero (a priori) probabilities of the results of such a measurement
belonging to disjoint Borel sets.

The irreversible and stochastic behaviour of the continuously observed
quantum system expressed by the so--called collapse or reduction of the wave
function has no analogue in the classical deterministic mechanics. The
Hamilton equations do not depend (for a nondemolition observation) on
whether the dynamical object is observed during its motion along its
trajectory. Ignoring that difference in the behaviour of classical and
quantum observed objects leads to various quantum paradoxes of Zeno kind
[5--11] which can be explained only in the way of a consistent investigation
of the disturbed stochastic dynamics of the quantum system undergoing an
observation.

It is quite natural to discuss this problem in the framework of stochastic
quantum mechanics of open systems [12, 13] on the basis of the theory of
nondemolition measurements developed recently [14--17]. The principle of a
nondemolition continuous observation of a quantum system can be formulated
as follows [17]:

{(i)} for any quantum measurement there exist observables $\hat{Q}(r),r\leq
t $, which commute for any $t$ with all Heisenberg operators $\hat{Z}(t)$ of
the system represented in the Hilbert space corresponding to ``the
system--measuring apparatus'',

{(ii)} according to the causality principle one does not impose any
conditions on the future observables $\hat{Q}(s),s>t$, with respect to the
past observables of the system $\hat{Z}(r),r\leq t$. A non--trivial
nondemolition observation in the above--mentioned sense is provided by
indirect measurements which can be only realized by considering the observed
quantum system as an open one.

From the experimental point of view it is natural to consider indirect
measurements because any measurement is taken with the help of some
experimental device. The indirect measurements allow to describe the state
changes resulting from the measurements of observables with continuous
spectra [4] which are assumed to be nonideal. The necessity to use indirect
measurements for the existence of the continual limit (with $\Delta
t\rightarrow 0$) for successive instantaneous measurements taken at instants
separated by $\Delta t$ is proved in Ref. [18].

In this paper we shall illustrate the approach of the continuous quantum
nondemolition measurement on the example of resolving the quantum Zeno
paradox for a three--dimensional free particle undergoing an observation
modeling the measurement of a trajectory of a quantum particle in a bubble
chamber briefly reported in Ref. [17] and for the one--dimensional case in
[19].

Sec. II has a preparatory character, we present here a stochastic model of a
continuous nondemolition observation of a quantum system interacting with $M$%
--dimensional Bose field reservoir modeling the measuring device, proposed
in [14--17].

In Sec. III we derive the filtering equation -- the stochastic differential
equation describing the time--development of the wave function of the
quantum system observed by means of the vector ``field coordinate'' process.
This equation was recently obtained with the help of quantum filtration
method [20, 21]. The presented derivation -- via stochastic instrument in
the sense of Davies and Lewis [2, 3] -- generalizes the result of Ref. [22]
to the case of multidimensional observation (the infinite dimensional and
general cases see in [23, 24]).

In Sec. IV we solve the filtering equation for the three--dimensional free
quantum particle undergoing the continuous nondemolition observation of its
position. We prove that the dispersion of the Gaussian wave packet does not
spread out in time but tends to the finite limit $\lim_{t\rightarrow \infty
}\tau ^{2}(t)=(\hbar /2\lambda m)^{1/2}$, where $m>0$ is the mass of the
observed particle and $\lambda $ stands for the accuracy coefficient of the
indirect nondemolition measurement of the particle position.

\section{Stochastic model of a continuous multidimensional diffusion
observation of a quantum system}

Let us assume that a quantum system $\mathcal{S}$ living in the Hilbert
space $\mathcal{H}_{0}$ is coupled at instant $t=0$ to the reservoir
(measuring device) consisting of $M$ independent Bose fields described by
vector--operators $\mathbf{b}(t)=[b_{j}(t)]_{1}^{M},\;\mathbf{b}%
^{+}(t)=[b_{j}^{+}(t)]_{1}^{M}$ acting in $\mathcal{F}=\mathcal{F}_{\text{sym%
}}(\mathbb{C}^{M}\otimes L^{2}(\mathbb{R}_{+}))$, the symmetric Fock space
over $\mathbb{C}^{M}\otimes L^{2}(\mathbb{R}_{+})$. The Bose field operators
satisfy the canonical commutation relations 
\begin{equation}
\lbrack b_{j}(t),\;b_{k}^{+}(s)]=\delta _{jk}\delta
(t-s)\;,\;[b_{j}(t),\;b_{k}(s)]=0(j,\;k=1,\dots ,\;M).  \tag{2.1}
\label{2.1}
\end{equation}%
The reservoir is assumed to be initially prepared in the vacuum state; $%
\langle b_{k}(t)\rangle _{v}=\langle b_{k}^{+}(t)\rangle _{v}=\langle
b_{k}^{+}(t)b_{k}(s)\rangle _{v}=0\;,\;\langle b_{k}(t)b_{l}^{+}(s)\rangle
_{v}=\delta _{kl}\delta (t-s)$. The real and imaginary parts of $\mathbf{b}%
(t)$ defined as $\func{Re}\mathbf{b}(t)={\frac{1}{2}}(\mathbf{b}(t)+\mathbf{b%
}^{+}(t)),\func{Im}\mathbf{b}(t)={\frac{1}{2\mathrm{i}}}(\mathbf{b}(t)-%
\mathbf{b}^{+}(t)$ do not commute, but each of them has the statistical
properties of the (classical) standard $M$--dimensional white noise.
Similarly as in the classical case [25] the time--evolution of the system
interacting with the reservoir can be described in a mathematically rigorous
way in terms of a stochastic differential equation [12, 13]. A quantum
stochastic calculus (QSC) of Ito type has been developed by Hudson and
Parthasarathy [12]. Here we give the formal rules of QSC which will be
needed in our paper.

Let us define annihilation and creation processes 
\begin{equation}
B_{j}(t)=\int_{0}^{t}b_{j}(s)\mathrm{d}s\;,\;B_{j}^{+}(t)=%
\int_{0}^{t}b_{j}^{+}(s)\mathrm{d}s\;,  \tag{2.2}  \label{2.2}
\end{equation}%
which satisfy the following commutation relations 
\begin{equation}
\lbrack B_{j}(t)\;,\;B_{k}^{+}(s)]=\delta _{jk}\min
(t,s)\;,\;[B_{j}(t)\;,\;B_{k}(s)]=0.  \tag{2.3}
\end{equation}%
The pair $\mathbf{B}(t)=[B_{j}(t)]_{1}^{M}\;,\;\mathbf{B}%
^{+}(t)=[B_{j}^{+}(t)]_{1}^{M}$ is the quantum analogue of standard complex $%
M$--dimensional Wiener diffusion process. The stochastic differentials of
the processes in (2.2) 
\begin{equation*}
\mathrm{d}B_{j}(t)=B_{j}(t+\mathrm{d}t)-B(t)\;,\;\mathrm{d}%
B_{j}^{+}(t)=B_{j}^{+}(t+\mathrm{d}t)-B_{j}^{+}(t)
\end{equation*}%
satisfy the multiplication rules 
\begin{equation}
\mathrm{d}B_{j}(t)\mathrm{d}B_{k}^{+}(t)=\delta _{jk}\mathrm{d}t  \tag{2.4}
\end{equation}%
and all other products involving $\mathrm{d}B_{j}(t)\;,\;\mathrm{d}%
B_{j}^{+}(t)$ and $\mathrm{d}t$ are equal to zero [12]. The
Hudson--Parthasarathy differentiation formula [12] for the product $M(t)N(t)$
of the adapted processes (the operators on $\mathcal{H}_{0}\otimes \mathcal{F%
}$ which depend on $\mathbf{B}(s)$ and $\mathbf{B}^{+}(s)$ only for times $%
s\leq t$) reads 
\begin{equation}
\mathrm{d}(M(t)\cdot N(t))=\mathrm{d}M(t)\cdot N(t)+M(t)\cdot \mathrm{d}N(t)+%
\mathrm{d}M(t)\cdot \mathrm{d}N(t).  \tag{2.5}
\end{equation}

We shall assume a unitary time--evolution of the compound quantum system in $%
\mathcal{H}_{0}\otimes \mathcal{F}$. The unitary evolution operator $U(t)$
for the system $\mathcal{S}$ coupled to the Bose reservoir is assumed to
satisfy the Ito quantum stochastic differential equation (QSDE) in the form
[12, 13] 
\begin{equation}
\mathrm{d}U(t)=\left[ \sum_{j}(L_{j}\mathrm{d}B_{j}^{+}(t)-L_{j}^{+}\mathrm{d%
}B_{j}(t))-K\mathrm{d}t\right] U(t)\;,\;U(0)=I,  \tag{2.6}
\end{equation}%
where 
\begin{equation}
K={\frac{\mathrm{i}}{\hbar }}H+{\frac{1}{2}}\sum_{j}L_{j}^{+}L_{j}. 
\tag{2.7}
\end{equation}%
In these formulas $H$ stands for the Hamiltonian of $\mathcal{S}$, $\mathrm{i%
}\hbar \sum (L_{j}\mathrm{d}B_{j}^{+}-L_{j}^{+}\mathrm{d}B_{j})$ describes
the interaction between $\mathcal{S}$ and the fields, $-{\frac{1}{2}}\sum
L_{J}^{+}L_{j}$ is the Ito correction term. (If one applied instead of (2.6)
a QSDE based on the quantum Stratonovich integral [13] this term would
disappear). With the help of (2.6) the Heisenberg equation of motion for any
observable of $\mathcal{S}$ can be easily obtained. By applying to the
product 
\begin{equation}
\hat{Z}(t)=U^{+}(t)ZU(t)  \tag{2.8}  \label{2.8}
\end{equation}%
the quantum Ito formula (2.5), Eq. (2.6) and its adjoint equation, one can
check with the help of (2.4) and (2.7) that the Heisenberg observable $\hat{Z%
}(t)$ satisfies the following QSDE 
\begin{equation}
\mathrm{d}\hat{Z}+\left( \hat{K}^{+}\hat{Z}+\hat{Z}\hat{K}-\sum_{j}\hat{L}%
_{j}^{+}\hat{Z}\hat{L}_{j}\right) \mathrm{d}t=\sum_{j}\left( \left[ \hat{Z},%
\hat{L}_{j}\right] \mathrm{d}B_{j}^{+}+\left[ \hat{L}_{j}^{+},\hat{Z}\right] 
\mathrm{d}B_{j}\right) ,  \tag{2.9}  \label{2.9}
\end{equation}%
where we have employed the simplified notation: $\hat{Z}$ for $\hat{Z}(t)$
etc.

Eq. (2.6) or Eqs. (2.9) describe the distorted dynamics of the initially
closed quantum system $\mathcal{S}$ under the stochastic interaction with
the Bose fields. The fields, however, do not only disturb the system. They
also give some possibility of a continuous (in time) observation of $%
\mathcal{S}$. Let us first pay attention to their time--development. In the
Heisenberg picture, the processes 
\begin{equation}
\hat{B}_{j}(t)=U^{+}(t)B_{j}(t)U(t)  \tag{2.10}
\end{equation}%
exhibit a useful property [26]: they remain unchanged for all times $s\geq t$%
, i.e. 
\begin{equation}
\hat{B}_{j}(t)=U^{+}(s)B_{j}(t)U(s)\;,\;s\geq t.  \tag{2.11}
\end{equation}%
Obviously, the same holds for the creation process $B^{+}(t)$. The property
(2.11) results essentially from two facts: Eq. (2.6) is written in the
interaction picture with respect to a free dynamics of the fields and the
coupling between $\mathcal{S}$ and fields is singular. The vector
annihilation and creation processes $\mathbf{B}(t)$ and $\mathbf{B}^{+}(t)$
are called input (annihilation, creation) processes while $\hat{\mathbf{B}}%
\;,\;\hat{\mathbf{B}}^{+}(t)$ are called output processes [13]. The input
processes describe Bose fields before their interaction with $\mathcal{S}$,
the output ones -- after the interaction. Note that due to (2.11) the output
processes satisfy the nondemolition conditions [17] 
\begin{equation}
\lbrack \hat{\mathbf{B}}(s),\hat{Z}(t)]=U^{+}(t)[\mathbf{B}(s),Z]U(t)=0\quad
\forall s\leq t  \tag{2.12}
\end{equation}

0Let us consider the continuous measurement of the output vector
\textquotedblleft field coordinate\textquotedblright\ (\textquotedblleft
diffusion\textquotedblright ) process 
\begin{equation*}
\hat{\mathbf{Q}}(t)=\hat{\mathbf{B}}(t)+\hat{\text{$\mathbf{B}$}}%
^{+}(t)=U^{+}(t){\mathbf{Q}}(t)U(t),
\end{equation*}%
where ${\mathbf{Q}}(t)=\mathbf{B}(t)+\mathbf{B}^{+}(t)$ is the input vector
Wiener process. From (2.3) it follows that 
\begin{equation}
\lbrack \hat{\mathbf{Q}}(t),\hat{\mathbf{Q}}(t^{\prime })]=0\quad
Vt,t^{\prime }\geq 0  \tag{2.13}
\end{equation}%
i.e. the output Hermitian process $\hat{\mathbf{Q}}$ is selfnondemolition.
Therefore the output process $\hat{\mathbf{Q}}(t)$ can be observed as a
classical process. Due to (2.12), (2.13) the measurement of $\hat{\mathbf{Q}}
$ is nondemolition [16, 17] with respect to the time--evolution of the
system, for any $Z$ 
\begin{equation}
\lbrack \hat{\mathbf{Q}}(s),\hat{Z}(t)]=0\quad Vs\leq t.  \tag{2.14}
\end{equation}%
This means that the measurement of ${\mathbf{Q}}$ disturbs neither the
present nor the future state of the system $\mathcal{S}$. The stochastic
differential equation for $\hat{\mathbf{Q}}(t)$, which can be easily
obtained in the same way as Eq. (2.9), has the form 
\begin{equation}
\mathrm{d}\hat{\mathbf{Q}}(t)=({{\mathbf{\hat{L}}}}(t)+{{\mathbf{\hat{L}}}}%
^{+}(t))\mathrm{d}t+\mathrm{d}{\mathbf{Q}}(t).  \tag{2.15}
\end{equation}%
Therefore, the process $\hat{\mathbf{Q}}(t)$ contains some information about 
$\mathcal{S}$.

Eq. (2.6) does not include any observation, it describes the perturbed
dynamics of the unobserved system $\mathcal{S}$ (represented in $\mathcal{H}%
_{0}\otimes \mathcal{F}$). Following Refs. 16, 17 we shall call it the prior
dynamics. Similarly, for any initial systematic observable $Z$, (2.9) is the
equation for the unobserved process $\hat{Z}(t)$. But for each $Z$ we have
the possibility of considering Eq. (2.9) together with Eq. (2.15),
consequently $\hat{Z}(t)$ for any initial $Z$ becomes partially observed. As
it is proved in Ref. [20] the condition (2.14) gives the possibility to
define the posterior (observed) mean values of $\hat{Z}(t)$ under the
condition of observation of any nonanticipating function of $\hat{\mathbf{Q}}
$ up to the moment $t$. It turns out [16, 17] that if the Bose reservoir is
initially prepared to be in the vacuum state and the initial state of $%
\mathcal{S}$ is pure, then the posterior state of $\mathcal{S}$ is a pure
one.

\section{Quantum filtering equation}

In this section we shall derive the quantum filtering equation -- the QSDE
which describes the time--development of the posterior state of the quantum
system $\mathcal{S}$ undergoing the $M$--dimensional diffusion observation
(2.15). It shall be done with the help of the method of solving the
differential equation for the generating map of the corresponding instrument
[2, 3]. For $M=1,\infty $ this approach was applied by one of us (V.P.B.) in
Ref. [22--24].

Let us denote by $\nu =\otimes _{j=1}^{M}\nu _{j}$ the standard product
Wiener probability measure on the space $\Omega $ of continuous trajectories 
${\mathbf{q}}=\{{\mathbf{q}}(t)|t>0\}$ of the observed process $\hat{\mathbf{%
Q}}$ restricted to the space $\Omega ^{t}=\{{\mathbf{q}}(t)|{\mathbf{z}}\in
\Omega \}$ of the trajectories stopped at $t:{\mathbf{q}}^{t}=\{{\mathbf{q}}%
(r)|r\leq t\}$. Consider the instrument $\mathcal{I}^{t}$ on the algebra of
operators $Z$ of the observed quantum system $\mathcal{S}$ as a function of
the observed event $\mathrm{d}\mathbf{q}$ up to the instant $t$. $\mathcal{I}%
^{t}$, by its definition, defines the time--evolution $\rho \mapsto \rho
^{t}(\mathrm{d}q)$ of an initial state functional $\rho :Z\mapsto \rho
\lbrack Z]$ of $\mathcal{S}$ to the state $\rho ^{t}(\mathrm{d}\mathbf{q}%
)=\rho \circ \mathcal{I}^{t}(\mathrm{d}\mathbf{q})$ normalized to the
probability $\mu ^{t}{(}\mathrm{d}{\mathbf{q})}=\rho \left[ \mathcal{I}^{t}(%
\mathrm{d}\mathbf{q})[I]\right] $.

Define the generating map of $\mathcal{I}^{t}$ in the following way (cf.
also Refs. [27, 28]) 
\begin{equation}
\Gamma ({\mathbf{l}},t)[Z]=\int_{\Omega ^{t}}\exp \left\{ \int_{0}^{t}{%
\mathbf{l}}(r)\mathrm{d}\mathbf{q}(r)\right\} \mathcal{I}^{t}(\mathrm{d}%
\mathbf{q})[Z],  \tag{3.1}
\end{equation}%
where ${\mathbf{l}}(t)=[l_{j}(t)]_{1}^{M}$ with components $l_{j}$ being
integrable $c$--valued functions. The generating map can also be defined by
the condition 
\begin{equation}
\langle \psi |\Gamma ({\mathbf{l}},t)[Z]\psi \rangle =\langle \hat{Y}({%
\mathbf{l}},t)\hat{Z}(t)\rangle ,  \tag{3.2}
\end{equation}%
where 
\begin{equation}
\hat{Y}({\mathbf{l}},t)=\exp \left\{ \sum_{j=1}^{M}\int_{0}^{t}l_{j}(r)%
\mathrm{d}\hat{Q}_{j}(r)\right\} =\dprod\limits_{j=1}^{M}\exp \left\{
\int_{0}^{t}l_{j}(r)\mathrm{d}\hat{Q}_{j}(r)\right\} .  \tag{3.3}
\end{equation}%
The mean value on the right hand side of (3.2) is taken with respect to $%
\psi \otimes \delta _{\phi }$ with $\psi \in \mathcal{H}_{0}$ being an
(arbitrary) initial pure state of $\mathcal{S}$ and $\delta _{\phi }\in 
\mathcal{F}$ the vacuum state vector for the fields. Note that the $M$%
--exponential output process $\hat{Y}({\mathbf{l}},t)$ given by (3.3) is
nondemolition and selfnondemolition.

Let us now find the differential equation for the generating map $\Gamma ({%
\mathbf{l}},t)$ of the instrument $\mathcal{I}^{t}$. According to (3.2) it
can be done by finding the differential equation for the mean value $\langle 
\hat{Y}({\mathbf{l}},t)\hat{Z}(t)\rangle $. First we obtain the stochastic
differential equation for $\hat{G}(t)=\hat{Y}(t)\hat{Z}(t)$. Let us write $%
\hat{G}(t)$ in the form $\hat{G}(t)=U^{+}(t)G(t)U(t)=U^{+}(t)Y(t)ZU(t)$,
where $Y(t)$ is the input process corresponding to (3.3): 
\begin{equation}
Y({\mathbf{l}},t)=\exp \left\{ \sum_{j=1}^{M}\int_{0}^{t}l_{j}(r)\mathrm{d}%
Q_{j}(r)\right\} .  \tag{3.4}
\end{equation}
Then from Ito's formula (2.5) applied to the product $\hat{G}=U^{+}GU$ we
get 
\begin{eqnarray*}
\mathrm{d}\hat{G} &=&U^{+}\left[ \sum_{j}\left( {\frac{1}{2}}%
l_{j}^{2}G+L_{j}^{+}Gl_{j}+l_{j}GL_{j}+L_{j}^{+}GL_{j}\right) -K^{+}G-GK%
\right] U\mathrm{d}t \\
&&+U^{+}\!\left[ \sum_{j}\left( L_{j}^{+}G+G(l_{j}-L_{j}^{+})\right) \mathrm{%
d}B_{j}+(GL_{j}+(l_{j}-L_{j})G)\mathrm{d}B_{j}^{+}\right] U,\ 
\end{eqnarray*}
where we have used (2.6), multiplication rules (2.4) and the stochastic
differential of $G,\mathrm{d}G=\mathrm{d}Y\cdot Z$ with 
\begin{equation}
\mathrm{d}Y({\mathbf{l}},t)=\sum_{j}\left( l_{j}(t)\mathrm{d}Q_{j}(t)+{\frac{%
1}{2}}l_{j}^{2}(t)\mathrm{d}t\right) Y({\mathbf{l}},t)  \tag{3.6}
\end{equation}
which can be obtained from (3.4) by classical Ito's formula [25].

Eq. (3.5) yields the following differential equation for the mean value of $%
\hat{G}(t)=\hat{Y}({\mathbf{l}},t)\hat{Z}(t)$ 
\begin{equation}
\langle \mathrm{d}\hat{G}\rangle =\langle \hat{\eta}(t)|\sum_{j}\bigl[{\frac{%
1}{2}}l_{j}^{2}G+l_{j}(L_{j}^{+}G+GL_{j})+L_{j}^{+}GL_{j}\bigr]-(K^{+}G+GK)|%
\hat{\eta}(t)\rangle \mathrm{d}t  \tag{3.7}
\end{equation}
with $\hat{\eta}(t)=U(t)\eta \;,\;\eta =\psi \otimes \delta _{\phi }$. Note
that the mean values of terms containing $\mathrm{d}B_{j}$ and $\mathrm{d}%
B_{j}^{+}$ in (3.5) do not appear in (3.7), they are equal to zero, because
for each $j$ 
\begin{equation}
\mathrm{d}B_{j}(t)U(t)\eta =U(t)\mathrm{d}B_{j}(t)\eta =0.  \tag{3.8}
\end{equation}
From (3.2) and (3.7) one can easily get the forward differential equation
for the generating map $\Gamma $: 
\begin{equation}
{\frac{\mathrm{d}}{\mathrm{d}t}}\Gamma \lbrack Z]=\Gamma \left[
\sum_{j}\left( {\frac{1}{2}}%
l_{j}^{2}Z+l_{j}(L_{j}^{+}Z+ZL_{j})+L_{j}^{+}ZL_{j}\right) -K^{+}Z-ZK\right]
\tag{3.9}
\end{equation}
with the initial condition $\Gamma ({\mathbf{l}},0)[Z]=Z$.

We shall prove that the solution of (3.9) has the form 
\begin{equation}
\Gamma ({\mathbf{l}},t)[Z]=\int_{\Omega ^{t}}Y({\mathbf{l}},{\mathbf{q}}%
^{t})V^{+}({\mathbf{q}}^{t})ZV({\mathbf{q}}^{t})\mathrm{d}\nu ({\mathbf{q}}%
^{t})  \tag{3.10}
\end{equation}
with the stochastic propagator $V(t)$ being the solution of a QSDE in the
form 
\begin{equation}
\mathrm{d}V(t)=-KV(t)\mathrm{d}t+\sum_{j}L_{j}\mathrm{d}Q_{j}(t)\;,\;V(0)=I.
\tag{3.11}
\end{equation}
Let us define the stochastic map $\Phi (t)$ from the algebra of observables
of $\mathcal{S}$ into itself 
\begin{equation}
\Phi (t)[Z]=V^{+}{(t)}ZV(t).  \tag{3.12}
\end{equation}
Then from Ito's formula (2.5) applied to the product appearing in (3.12) we
get 
\begin{equation*}
\mathrm{d}(\Phi (t)[Z])=\mathrm{d}V^{+}(t)ZV(t)+V^{+}(t)Z\mathrm{d}V(t)+%
\mathrm{d}V^{+}(t)Z\mathrm{d}V(t).
\end{equation*}
By making use of (3.11) we obtain the recursive filtering equation for the
stochastic map $\Phi (t)$ 
\begin{equation}
\mathrm{d}(\Phi (t)[Z])=\Phi (t)\left[ \sum_{j}L_{j}^{+}ZL_{j}-K^{+}Z-ZK%
\right] +\sum_{j}\Phi (t)[L_{j}^{+}Z+ZL_{j}]\mathrm{d}Q_{j}(t)\;  \tag{3.13}
\end{equation}
with $\Phi (0)[Z]=Z.$ The stochastic map (3.12) defines for any trajectory ${%
\mathbf{q}}$ the selective instrument, $\Phi (t)({\mathbf{q}})[Z]=\Phi ({%
\mathbf{q}}^{t})[Z]=V^{+}({\mathbf{q}}^{t})ZV({\mathbf{q}}^{t})$. Taking
into account that 
\begin{equation*}
\mathrm{d}(Y({\mathbf{l}},t)\Phi (t)[Z])=\mathrm{d}Y({\mathbf{l}},t)\Phi
(t)[Z]+Y({\mathbf{l}},t)\mathrm{d}\Phi (t)[Z]+\mathrm{d}Y({\mathbf{l}},t)%
\mathrm{d}\Phi (t)[Z]
\end{equation*}
\begin{equation*}
+Y({\mathbf{l}},t)\Phi (t)\bigl[\sum_{j}\bigl({\frac{1}{2}}%
l_{j}^{2}(t)Z+l_{j}(t)(L_{j}^{+}Z+ZL_{j})+L_{j}^{+}ZL_{j}\bigr)-K^{+}Z-ZK%
\bigr]\mathrm{d}t
\end{equation*}
and averaging it with respect to the standard product Wiener measure one
obtains (3.9) for the mean value (3.10) of the product $Y({\mathbf{l}},{%
\mathbf{q}}^{t})\Phi ({\mathbf{q}}^{t})[Z]$.

So, the wave function $\hat{\chi}(t)=V(t)\psi $ of the system $\mathcal{S}$
under the continuous nondemolition diffusion observation of $\hat{\mathbf{Q}}
$, satisfies the stochastic dissipative differential equation 
\begin{equation}
\mathrm{d}\hat{\chi}(t)+\left( {\frac{\mathrm{i}}{\hbar }}H+{\frac{1}{2}}%
\sum_{j}L_{j}^{+}L_{j}\right) \hat{\chi}(t)\mathrm{d}t=\sum_{j}L_{j}\hat{\chi%
}(t)\mathrm{d}Q_{j}(t),\;\hat{\chi}(0)=\psi .  \tag{3.14}
\end{equation}%
Eq. (3.14) plays an analogous role to the Schr\"{o}dinger equation for the
unobserved quantum system. (In (3.14) $\mathrm{d}{\mathbf{q}}$ can be
replaced with $\mathrm{d}\hat{\mathbf{Q}}$ because in the Schr\"{o}dinger
picture ${\mathbf{Q}}$ and $\hat{\mathbf{Q}}$ coincide.) The posterior wave
function $\hat{\chi}(t)$ is normalized to the probability density 
\begin{equation*}
p({\mathbf{q}}^{t})=\langle V({\mathbf{q}}^{t})\psi |V({\mathbf{q}}^{t})\psi
\rangle \equiv \hat{p}(t)({\mathbf{q}})
\end{equation*}%
of the observed process $\hat{\mathbf{Q}}$ with respect to the standard
product Wiener measure of the input process ${\mathbf{Q}}$. It follows from
the integral representation of (3.2) 
\begin{equation}
\langle \hat{Y}({\mathbf{l}},t)\hat{Z}(t)\rangle =\int_{\Omega ^{t}}Y({%
\mathbf{l}},q^{t})\langle V({\mathbf{q}}^{t})\psi |ZV({\mathbf{q}}%
^{t})\rangle \mathrm{d}\nu ({\mathbf{q}}^{t})  \tag{3.15}
\end{equation}%
giving for $Z=I$ the mean value of the output process (3.3) as the
generating function of the output probability measure 
\begin{equation*}
\mathrm{d}\mu ({\mathbf{q}}^{t})=p({\mathbf{q}}^{t})\mathrm{d}\nu ({\mathbf{q%
}}^{t}).
\end{equation*}%
The formula (3.15) defines the posterior mean value $\langle Z\rangle (q^{t})
$ as 
\begin{equation*}
\langle Z\rangle ({\mathbf{q}}^{t})=\langle \psi ({\mathbf{q}}^{t})|Z\psi ({%
\mathbf{q}}^{t})\rangle \equiv \hat{z}(t)({\mathbf{q}})
\end{equation*}%
in terms of the normalized posterior wave function $\hat{\psi}(t){\mathbf{q}}%
)=\psi ({\mathbf{q}}^{t})\psi ({\mathbf{q}}^{t})=\chi ({\mathbf{q}}%
^{t})/q(t)^{1/2}$.

The normalized posterior wave function $\hat{\psi}(t)$ satisfies the
nonlinear stochastic wave equation [16, 17] 
\begin{equation}
\mathrm{d}\hat{\psi}(t)+\left( {\frac{1}{2}}\sum_{j}\tilde{L}_{j}^{+}(t)%
\tilde{L}_{j}(t)+{\frac{\mathrm{i}}{\hbar }}\tilde{H}(t)\right) \hat{\psi}(t)%
\mathrm{d}t=\sum_{j}\tilde{L}_{j}(t)\mathrm{d}\tilde{Q}_{j}(t)\hat{\psi}(t),
\tag{3.16}
\end{equation}%
where 
\begin{equation*}
\tilde{L}_{j}(t)=L_{j}-\func{Re}\,\hat{l}_{j}(t)\;,\;\tilde{H}(t)=H-\hbar
\sum_{j}\func{Re}\,\hat{l}_{j}(t)\func{Im}L_{j},
\end{equation*}%
and $\mathrm{d}\tilde{Q}_{j}(t)=\mathrm{d}Q_{j}(t)-2\func{Re}\,\hat{l}_{j}(t)%
\mathrm{d}t$ is the so--called Wiener innovating process.

Eq. (3.16) can be obtained from Eq. (3.14) in the following way. Writing $%
\hat{\psi}(t)$ in the form $\hat{\psi}(t)=\hat{\chi}(t)(\hat{\chi}^{+}(t)%
\hat{\chi}(t))^{-1/2}$ we get 
\begin{equation}
\mathrm{d}\hat{\psi}=\mathrm{d}\hat{\chi}\cdot (\hat{\chi}^{+}\hat{\chi}%
)^{-1/2}+\hat{\chi}\cdot \mathrm{d}[(\hat{\chi}^{+}\hat{\chi})^{-1/2}]+%
\mathrm{d}\hat{\chi}\cdot \mathrm{d}[(\hat{\chi}^{+}\hat{\chi})^{-1/2}]\ . 
\tag{3.17}
\end{equation}
For $\hat{\chi}$ satisfying Eq. (3.14) one finds easily 
\begin{equation*}
\mathrm{d}(\hat{\chi}^{+}\hat{\chi})=2\sum_{j}\hat{\chi}^{+}(\func{Re}%
\,L_{j})\hat{\chi}\mathrm{d}Q_{j}
\end{equation*}
and by the classical Ito formula 
\begin{equation}
\mathrm{d}[(\hat{\chi}^{+}\hat{\chi})^{-1/2}]=(\hat{\chi}^{+}\hat{\chi}%
)^{-1/2}\left\{ -\sum_{j}\func{Re}\,\hat{l}_{j}(t)\mathrm{d}Q_{j}+{\frac{3}{2%
}}\sum_{j}(\func{Re}\,\hat{l}_{j}(t))^{2}\mathrm{d}t\right\}  \tag{3.18}
\end{equation}
\nopagebreak
Finally combining (3.17), (3.18) and (3.14) yields Eq. (3.16).

\section{Watchdog effect}

The Schr\"{o}dinger equation for a free particle 
\begin{equation}
\dot{\psi}-{\frac{\mathrm{i}\hbar }{2m}}\Delta \psi =0  \tag{4.1}
\end{equation}%
describes the effect of spreading out of the wave packet. The probability of
detection of the quantum particle in any finite coordinate region tends to
zero as time increases.

On the other hand experimental data on observed quantum particles show
well--localized paths of quantum particles (for instance in bubble chamber
experiments). This phenomenon being an example of the watchdog effect is
also known as quantum Zeno paradox [5] because it is in contradiction with
predictions of Eq. (4.17). The above paradox of the orthodox quantum
mechanics can be resolved in the framework of posterior quantum dynamics for
observed quantum systems by using nondemolition filtering methods.

The typical observations in quantum systems are indirect (in the bubble
chamber the path of an ionizing particle is made visible by a string of
vapor bubbles), moreover one has to consider the interaction with the
measuring device, hence the observed quantum object should be considered as
an open quantum system.

The aim of this section is to demonstrate the watchdog effect that occurs
for a free quantum particle coupled to the three--dimensional Bose field in
the vacuum state (measuring device), the position of which is continuously
observed. We shall consider an indirect measurement of the particle position 
${\mathbf{X}}=[X_{1},X_{2},X_{3}]$ therefore we choose the coupling operator 
${\mathbf{L}}$ (cf. (2.6) and (2.15)) to be proportional to ${\mathbf{X}}$, 
\begin{equation}
{\mathbf{L}}=(\lambda /2)^{1/2}{\mathbf{X}}.  \tag{4.2}
\end{equation}%
With such a choice of ${\mathbf{L}}$ we get the QSDEs describing the
perturbed dynamics of the particle in the Heisenberg picture by putting for $%
Z$ in Eq. (2.9) the position and momentum components 
\begin{equation}
\mathrm{d}{{\mathbf{\hat{X}}}}(t)={\frac{1}{m}}{{\mathbf{\hat{P}}}}(t)%
\mathrm{d}t\;,\;\mathrm{d}{{\mathbf{\hat{P}}}}(t)=(2\lambda )^{1/2}\hbar \ 
\mathrm{d}(\func{Im}\text{$\mathbf{B}$}^{+}(t)).  \tag{4.3}
\end{equation}%
Eqs. (4.3) describe the motion of the particle upon the stochastic
(Langevin) force $f(t)=(2\lambda )^{1/2}\hbar \ \func{Im}\mathbf{b}%
^{+}(t)=-(2\lambda )^{1/2}\hbar \ \func{Im}\mathbf{b}(t)(cf.(2.2))$ from the
Bose reservoir.

The observed nondemolition field coordinate process $\hat{\mathbf{Q}}(t)$
((2.13)) satisfies due to (2.15) and (4.2) the QSDE in the form 
\begin{equation}
\mathrm{d}\hat{\mathbf{Q}}(t)=(2\lambda )^{1/2}{{\mathbf{\hat{X}}}}(t)%
\mathrm{d}t+\mathrm{d}{\mathbf{Q}}(t).  \tag{4.4}
\end{equation}

Eq. (4.4) describes the indirect (and imperfect) measurement of the particle
position. Note that in terms of generalized derivatives of the processes $%
\hat{\mathbf{Q}}$ and ${\mathbf{Q}}$ Eq. (4.4) can be written as $\dot{\hat{%
\mathbf{Q}}}(t)=(2\lambda )^{1/2}{{\mathbf{\hat{X}}}}(t)+2\func{Re}{{\overset%
{\cdot }{\mathbf{B}}}}(t)=(2\lambda )^{1/2}{{\mathbf{\hat{X}}}}(t)+2\func{Re}%
\mathbf{b}(t)$, therefore the (generalized) stochastic process $\dot{\hat{%
\mathbf{Q}}}(t)$ describes the measurement of ${{\mathbf{\hat{X}}}}(t)$
together with a random error given by the standard vector white noise $2%
\func{Re}\mathbf{b}(t)$. From the last formula one can see that the positive
constant $\lambda $ can be interpreted as the measurement accuracy
coefficient.

Let us denote by $\hat{\mathbf{q}}(t)=[\hat{q}_{j}(t)]_{j=1}^{3}$ and $\hat{%
\mathbf{q}}(t)=[\hat{q}_{j}(t)]_{j=1}^{3}$ the posterior mean values of
position and momentum of the observed particle. We have 
\begin{equation}
\hat{\mathbf{q}}(t)=\int \hat{\psi}(t,{\mathbf{x}})^{\ast }{\mathbf{x}}\hat{%
\psi}(t,{\mathbf{x}})\mathrm{d}{\mathbf{x}}\;,\;{{\mathbf{\hat{p}}}}(t)=\int 
\hat{\psi}(t,{\mathbf{x}})^{\ast }{\frac{\hbar }{\mathrm{i}}}\nabla \hat{\psi%
}(t,{\mathbf{x}})\mathrm{d}{\mathbf{x}}.  \tag{4.5}
\end{equation}%
According to (3.16) the posterior (normalized) wave function satisfies in
the considered case the stochastic wave equation which in the coordinate
representation has the form 
\begin{equation}
\mathrm{d}\hat{\psi}-\left( {\frac{\mathrm{i}\hbar }{2m}}\Delta \hat{\psi}-{%
\frac{\lambda }{4}}({\mathbf{x}}-\hat{\mathbf{q}})^{2}\hat{\psi}\right) 
\mathrm{d}t=\hat{\psi}\left( {\frac{\lambda }{2}}\right) ^{1/2}({\mathbf{x}}-%
\hat{\mathbf{q}})\mathrm{d}{{\mathbf{\tilde{Q}}}}\;,\;\hat{\psi}(0)=\psi  
\tag{4.6}
\end{equation}%
with 
\begin{equation*}
\mathrm{d}{{\mathbf{\tilde{Q}}}}(t)=\mathrm{d}{\mathbf{Q}}(t)-\left(
2\lambda \right) ^{1/2}\hat{\mathbf{q}}(t)\mathrm{d}t.
\end{equation*}

Let us now discuss the time--development of the posterior wave function
assuming that the initial state $\psi $ has the form of the Gaussian wave
packet, 
\begin{equation}
\psi ({\mathbf{x}})=(2\sigma _{q}^{2}\pi )^{-3/4}\exp \left\{ -{\frac{1}{%
4\sigma _{q}^{2}}}({\mathbf{x}}-{\mathbf{q}})^{2}+{\frac{\mathrm{i}}{\hbar }}%
{\mathbf{q}}{\mathbf{x}}\right\} \;,  \tag{4.7}
\end{equation}%
$\mathbf{p}$ and ${\mathbf{q}}$ denote the initial mean values of position
and momentum of the particle and $\sigma _{q}^{2}$ stands for the initial
dispersion of the wave packet. We shall prove that the solution of Eq. (4.6)
corresponding to the initial condition (4.7) has the form of Gaussian packet 
\begin{equation}
\hat{\psi}(t,{\mathbf{x}})=\hat{c}(t)\exp \left\{ -{\frac{1}{2}}\omega (t)({%
\mathbf{x}}-\hat{\mathbf{q}}(t))^{2}+{\frac{\mathrm{i}}{\hbar }}{{\mathbf{%
\hat{p}}}}(t){\mathbf{x}}\right\}   \tag{4.8}
\end{equation}%
with posterior mean values $\hat{\mathbf{q}}(t),{{\mathbf{\hat{p}}}}(t)$,
cf. (4.5), fulfilling linear filtration equations and $\omega (t)$
satisfying the Riccatti differential equation. In Eq. (4.8) $\hat{c}%
(t)=(2\tau _{q}^{2}\pi )^{-3/4}$ up to unessential stochastic phase factor
and $\tau _{q}^{2}=\hat{{\mathbf{q}}^{2}}-\hat{\mathbf{q}}^{2}$ is the
posterior position dispersion.

It is convenient to rewrite Eq. (4.6) in terms of the complex osmotic
velocity. By introducing%
\begin{equation*}
T(t,{\mathbf{x}})=R(t,{\mathbf{x}})+\mathrm{i}S(t,{\mathbf{x}})=\hbar \ ln%
\hat{\psi}(t,{\mathbf{x}}),
\end{equation*}%
next by Ito's rule%
\begin{equation*}
\mathrm{d}T(\hat{\psi})=T^{\prime }(\hat{\psi})\mathrm{d}\hat{\psi}+{\frac{1%
}{2}}T^{\prime \prime }(\hat{\psi})(\mathrm{d}\hat{\psi})^{2}
\end{equation*}%
applied to the function $T=\hbar \ln \psi \ $ and by taking into account that%
\begin{equation*}
(\mathrm{d}\hat{\psi})^{2}={\frac{\lambda }{2}}({\mathbf{x}}-\hat{\mathbf{q}}%
)^{2}\hat{\psi}^{2}\mathrm{d}t
\end{equation*}%
we obtain Eq. (4.6) in terms of $T$. From this equation we get for the
complex osmotic velocity 
\begin{equation*}
\mathbf{W}(t,{\mathbf{x}})={\frac{1}{m}}\nabla T(t,{\mathbf{x}})=\mathbf{U}%
(t,{\mathbf{x}})+\mathrm{i}\mathbf{V}(t,{\mathbf{x}})
\end{equation*}%
the following equation 
\begin{equation}
\mathrm{d}\mathbf{W}+\left[ {\frac{\hbar \lambda }{m}}({\mathbf{x}}-\hat{%
\mathbf{q}})-{\frac{\mathrm{i}}{2}}(\nabla \mathbf{W}^{2}+{\frac{\hbar }{m}}%
\Delta \mathbf{W})\right] \mathrm{d}t=\left( {\frac{\lambda }{2}}\right)
^{1/2}{\frac{\hbar }{m}}\mathrm{d}\mathbf{\tilde{Q}}.  \tag{4.9}
\end{equation}%
We shall look for the solution of Eq. (4.9) corresponding to the initial
condition 
\begin{equation}
\mathbf{W}(0,{\mathbf{x}})={\frac{\hbar }{m}}\nabla \ln \psi ({\mathbf{x}})={%
\frac{\hbar }{2m\sigma _{q}^{2}}}({\mathbf{q}}-{\mathbf{x}})+{\frac{\mathrm{i%
}}{m}}\mathbf{p}  \tag{4.10}
\end{equation}%
in the linear form 
\begin{equation}
\mathbf{W}(t,x)=\mathbf{\hat{W}}(t)-{\frac{\hbar }{m}}\omega (t){\mathbf{x}}
\tag{4.11}
\end{equation}%
where in accordance with (4.8) 
\begin{equation}
\mathbf{\hat{w}}(t)={\frac{\hbar }{m}}\omega (t)\mathbf{\hat{q}}(t)+{\frac{%
\mathrm{i}}{m}}\mathbf{\hat{p}}(t).  \tag{4.12}
\end{equation}%
By putting $\nabla \mathbf{W}^{2}=-{\frac{2\hbar \omega }{m}}\mathbf{W}$, $%
\Delta \mathbf{W}=0$ into (4.9) we obtain the following system of equations
for coefficients $\mathbf{\hat{w}}(t)$ and $\omega (t)$ 
\begin{equation}
\mathrm{d}{\mathbf{\hat{w}}}(t)+{\frac{\mathrm{i}\hbar }{m}}\omega (t)%
\mathbf{\hat{w}}(t)\mathrm{d}t=\left( {\frac{\lambda }{2}}\right) ^{1/2}{%
\frac{\hbar }{m}}\mathrm{d}{\mathbf{Q}}(t),\;\;\mathbf{\hat{w}}(0)={\frac{%
\hbar }{2m\sigma _{q}^{2}}}{\mathbf{q}}+{\frac{\mathrm{i}}{m}}\mathbf{p}, 
\tag{4.13}
\end{equation}%
\begin{equation}
{\frac{\mathrm{d}}{\mathrm{d}t}}\omega (t)+{\frac{\mathrm{i}\hbar }{m}}%
\omega (t)^{2}=\lambda ,\;\;\omega (0)={\frac{1}{2\sigma _{q}^{2}}}\;, 
\tag{4.14}
\end{equation}%
which define the solution of Eq. (4.9) in the form (4.11). From (4.12) we
get $\mathbf{\hat{q}}(t)=m\,\func{Re}\,\mathbf{\hat{w}}(t)/\hbar \func{Re}%
\,\omega (t)$ which is the root of the equation $\pmb\nabla R(t,{\mathbf{x}}%
)=m\mathbf{U}(t,{\mathbf{x}})=0$ for which the maximum of the posterior
density $|\hat{\psi}(t,{\mathbf{x}})|^{2}=\exp \left\{ {\frac{2}{\hbar }}R(t,%
{\mathbf{x}})\right\} $ is attained. The posterior mean value of momentum $%
\hat{p}(t)$ coincides with 
\begin{equation*}
m\mathbf{V}(t,\mathbf{\hat{q}}(t))=\nabla S(t,{\mathbf{x}})_{|{\mathbf{x}}=%
\mathbf{\hat{q}}(t)}\;\text{and by (4.12)}\;\mathbf{\hat{p}}(t)=\func{Im}(m%
\mathbf{\hat{w}}(t)-\hbar \omega (t)\mathbf{\hat{q}}(t)).
\end{equation*}

Eq. (4.12) gives the time--development of posterior mean values of position
and momentum, with the help of (4.13) and (4.14) we obtain the
Hamilton--Langevin equations 
\begin{eqnarray}
\mathrm{d}\mathbf{\hat{q}}(t)-{\frac{1}{m}}\mathbf{\hat{p}}(t)\mathrm{d}t &=&%
{\frac{(\lambda /2)^{1/2}}{\func{Re}\ \omega (t)}}\mathrm{d}\mathbf{\tilde{Q}%
}\left( t\right) ,\quad \mathbf{\hat{q}}(0)={\mathbf{q}},  \TCItag{4.15} \\
\cr\mathrm{d}\mathbf{\hat{p}}(t) &=&-\hbar {\frac{(\lambda /2)^{1/2}\func{Im}%
\omega (t)}{\func{Re}\ \omega (t)}}\mathrm{d}\mathbf{\tilde{Q}}(t),\quad 
\mathbf{\hat{p}}(0)=\mathbf{p}.\cr  \notag
\end{eqnarray}%
They are classical stochastic equations describing continuously and
indirectly observed position and momentum of a free quantum particle
disturbed by the measuring device (in the mean $\mathbf{\hat{p}}(t)$ and $%
\mathbf{\hat{q}}(t)$ coincide with $\mathbf{p}(t)=\mathbf{p}\;,\;{\mathbf{q}}%
(t)=\mathbf{p}t/m$).

One can check easily that for the posterior wave function in the form (4.8),
posterior momentum and position dispersions are given by formulas 
\begin{equation}
\tau _{q}^{2}(t)=1/2\func{Re}\,\omega (t),\;\tau _{p}^{2}(t)=\hbar
^{2}|\omega (t)|^{2}/2\func{Re}\ \omega (t)  \tag{4.16}
\end{equation}
with $\omega (t)$ being the solution of Eq. (4.14). These formulas yield the
Heisenberg inequality $\tau _{q}^{2}\tau _{p}^{2}\geq \hbar ^{2}/4$.

The general solution of Eq. (4.14) has the form 
\begin{equation}
\omega (t)=\alpha {\frac{\omega (0)+\alpha \tanh ({\frac{\lambda }{\alpha }}%
t)}{\omega (0)\tanh ({\frac{\lambda }{\alpha }}t)+\alpha }}\;,\;\alpha
=\left( {\frac{\lambda m}{2\hbar }}\right) ^{1/2}(1-\mathrm{i}).  \tag{4.17}
\end{equation}%
Obviously, $\lim_{t\rightarrow \infty }\omega (t)=\alpha $, i.e. $\alpha $
is the asymptotic stationary solution of Eq. (4.14). Consequently, the
posterior dispersions of position and momentum tend to finite limits
independent of its initial values 
\begin{equation}
\tau _{q}^{2}(\infty )=(\hbar /2\lambda m)^{1/2},\;\tau _{p}^{2}(\infty
)=\hbar (\lambda m\hbar /2)^{1/2}  \tag{4.18}
\end{equation}%
giving the localization of the observed quantum particle. As it follows from
(4.18) the asymptotic localization of the particle in the coordinate
representation is inversely proportional to its mass and the measurement
accuracy $\lambda $. It means that the particle of mass zero cannot be
localized by any measurement and heavy particles ($m\rightarrow \infty $)
can be localized at a point. Note that according to the dimension of $%
\lambda $, $[\lambda ]=(m^{2}\sec )^{-1}$, the measurement accuracy
coefficient can be interpreted as inversely proportional to the scattering
cross--section and characteristic time of transition process in a bubble
chamber.

\section*{References}

{[1]} J. von Neumann, \textit{Mathematical Foundations of Quantum Mechanics}
(Princeton Unive, Press, Princeton 1955).

{[2]} E.B. Davies, \textit{Quantum Theory of Open Systems} (Academic, London
1976).

{[3]} E.B. Davies and J.T. Lewis, Commun. Math. Phys. \textbf{17}, 239
(1970).

{[4]} M. Ozawa, J. Math. Phys. \textbf{25}, 79 (1984).

{[5]} B. Misra and E.C.G. Sudarshan, J. Math. Phys. \textbf{18}, 756 (1973).

{[6]} C.B. Chiu and E.C.G. Sudarshan, Phys. Rev. D \textbf{16}, 520 (1973).

{[7]} A. Peres, Am. J. Phys. \textbf{48}, 931 (1980).

{[8]} K. Kraus, Found. Phys. \textbf{11}, 547 (1981).

{[9]} C.B. Chiu, B. Misra and E.C.G. Sudarshan, Phys. Lett. \textbf{117}, 34
(1982).

{[10]} E. Joos, Phys. Rev. D \textbf{29}, 1626 (1984).

{[11]} D. Home and M.A.B. Whitaker, J. Phys. A: Math. Gen. \textbf{19}, 1847
(1986), \textbf{20}, 3339 (1987).

{[12]} R.L. Hudson and K.R. Parthasarathy, Commun. Math. Phys. \textbf{93},
301 (1984).

{[13]} C.W. Gardiner and M.J. Collett, Phys. Rev. A \textbf{31}, 3761 (1985).

{[14]} V.P. Belavkin, \textit{Radiotechnika i Elektronika} \textbf{25}, 1445
(1980).

{[15]} V.P. Belavkin, in \textit{Information Complexity and Control in
Quantum Physics}, edited by A. Blaquiere, S. Diner and G. Lochak (Springer,
Wien -- New York 1987), p. 311.

{[16]} V.P. Belavkin, in \textit{Stochastic Methods in Mathematics and
Physics}, edited by R. Gielerak and W. Karwowski (World Scientific,
Singapore 1989), p. 310.

{[17]} V.P. Belavkin, in \textit{Modeling and Control of Systems in
Engineering, Quantum Mechanics, Economics, and Biosciences}, ed. by A.
Blaquiere (Springer, Berlin 1989), p. 245.

{[18]} A. Barchielli, L. Lanz and G.M. Prosperi, Found. Phys. \textbf{13},
779 (1973).

{[19]} V.P. Belavkin and P. Staszewski, Phys. Lett. A \textbf{140}, 359
(1989).

{[20]} V.P. Belavkin, \textit{Quantum Stochastic Calculus and Quantum
Nonlinear Filtering}, Preprint Centro Matematico V. Volterra, Univ. di Roma
II n. 6 (1989).

{[21]} V.P. Belavkin, Phys. Lett. A \textbf{140}, 355 (1989).

{[22]} V.P. Belavkin, \textit{A Posterior Schr\"odinger Equation for a
Continuous Nondemolition Measurement}, J. Math. Phys., submitted.

{[23]} V.P. Belavkin, \textit{A continuous diffusion observation and
posterior quantum dynamics}, submitted to J. of Phys. A, Preprint N$^\circ$
Phys. Dep. University of Roma ``La Sapienza'', 1989.

{[24]} V.P. Belavkin, \textit{A field stochastic calculus and quantum
filtering in Fock space}. Preprint N$^\circ$ of Phys. Dep. University of
Rome ``La Sapienza'', 1989.

{[25]} C.W. Gardiner, \textit{Handbook of Stochastic Methods} (Springer,
Berlin 1983).

{[26]} A. Barchielli, in \textit{Quantum Probability and Applications III},
edited by L. Accardi and W. von Waldenfelds (Springer, Berlin 1988), p. 37.

{[27]} A. Barchielli, Phys. Rev. D \textbf{32}, 347 (1985).

{[28]} A. Barchielli, Phys. Rev. A \textbf{34}, 1642 (1986).

\end{document}